\documentclass[preprint2,tighten]{aastex6}
\pdfoutput=1 
\usepackage{amsmath,amstext}
\usepackage[T1]{fontenc}
\usepackage{apjfonts} 
\usepackage[figure,figure*]{hypcap}
\usepackage{color}

\usepackage{hyperref}
\usepackage{breakurl}
\shorttitle{BAR5}
\shortauthors{Wisniewski et al.}

\begin{document}
\title{High Fidelity Imaging of the Inner AU Mic Debris Disk: Evidence of Differential Wind Sculpting?}
\author{John P. Wisniewski\altaffilmark{1}}
\author{Adam F. Kowalski \altaffilmark{2,3}}
\author{James R.A. Davenport \altaffilmark{4}}
\author{Glenn Schneider\altaffilmark{6}}
\author{Carol A. Grady\altaffilmark{5}}
\author{Leslie Hebb\altaffilmark{7}}
\author{Kellen D. Lawson\altaffilmark{1}}
\author{Jean-Charles Augereau\altaffilmark{8}}
\author{Anthony Boccaletti\altaffilmark{9}}
\author{Alexander Brown\altaffilmark{16}}
\author{John H. Debes\altaffilmark{10}}
\author{Andras Gaspar\altaffilmark{6}}
\author{Thomas K. Henning\altaffilmark{11}}
\author{Dean C. Hines\altaffilmark{10}}
\author{Marc J. Kuchner\altaffilmark{12}}
\author{Anne-Marie Lagrange\altaffilmark{13}}
\author{Julien Milli\altaffilmark{14}}
\author{Elie Sezestre\altaffilmark{8}}
\author{Christopher C. Stark\altaffilmark{10}}
\author{Christian Thalmann\altaffilmark{15}}

\altaffiltext{1}{Homer L. Dodge Department of Physics and Astronomy, University of Oklahoma, 440 W. Brooks Street, Norman, OK 73019, USA; wisniewski@ou.edu}
\altaffiltext{2}{Department of Astrophysical and Planetary Sciences, University of Colorado Boulder, 2000 Colorado Ave, Boulder, CO 80305, USA}
\altaffiltext{3}{National Solar Observatory, University of Colorado Boulder, 3665 Discovery Drive, Boulder, CO 80303, USA}
\altaffiltext{4}{Department of Astronomy, University of Washington, Seattle, WA 98195, USA}
\altaffiltext{5}{Eureka Scientific, 2452 Delmer, Suite 100, Oakland, CA 96002 USA}
\altaffiltext{6}{Steward Observatory and the Department of Astronomy, The University of Arizona, 933 North Cherry Avenue, Tucson, AZ 85721 USA}
\altaffiltext{7}{Department of Physics, Hobart and William Smith Colleges, 300 Pulteney Street, Geneva, NY 14456, USA}
\altaffiltext{8}{Univ. Grenoble Alpes, CNRS, IPAG, 38000 Grenoble, France}
\altaffiltext{9}{LESIA, Observatoire de Paris, Université PSL, CNRS, Sarbonne Université, Univ. Paris Diderot, Sorbonne Paris Cité, 5 Place Jules Janssen, 92195 Meudon, France}
\altaffiltext{10}{Space Telescope Science Institute, Baltimore, MD 21218, USA}
\altaffiltext{11}{Max-Planck-Institut für Astronomie, Königstuhl 17, 69117 Heidelberg, Germany}
\altaffiltext{12}{Exoplanets and Stellar Astrophysics Laboratory, NASA Goddard Space Flight Center, 8800 Greenbelt Road, Greenbelt, MD 20771, USA}
\altaffiltext{13}{Université Grenoble Alpes, CNRS, IPAG, 38000, Grenoble, France}
\altaffiltext{14}{European Southern Observatory (ESO), Alonso de Córdova 3107, Vitacura, Casilla 19001, Santiago, Chile}
\altaffiltext{15}{Eidgenossiche Technische Hochschule}
\altaffiltext{16}{Center for Astrophysics and Space Astronomy, 389 UCB, University of Colorado,
Boulder, CO 80309-0389}

\begin{abstract}

We present new high fidelity optical coronagraphic imagery of the inner $\sim$50 au of AU Mic's edge-on debris disk using the BAR5 occulter of the Hubble Space Telescope Imaging Spectrograph (HST/STIS) obtained on 26-27 July 2018.  This new imagery reveals that ``feature A'', residing at a projected stellocentric separation of 14.2 au on SE-side of the disk, exhibits an apparent ``loop-like'' morphology at the time of our observations. The loop has a projected width of 1.5 au and rises 2.3 au above the disk midplane. We also explored TESS photometric observations of AU Mic that are consistent with evidence of two starspot complexes in the system. The likely co-alignment of the stellar and disk rotational axes breaks degeneracies in detailed spot modeling, indicating that AU Mic's projected magnetic field axis is offset from its rotational axis. We speculate that small grains in AU Mic's disk could be sculpted by a time-dependent wind that is influenced by this offset magnetic field axis, analogous to co-rotating Solar interaction regions that sculpt and influence the inner and outer regions of our own Heliosphere. Alternatively, if the observed spot modulation is indicative of a significant mis-alignment of the stellar and disk rotational axes, we suggest the disk could still be sculpted by the differential equatorial versus polar wind that it sees with every stellar rotation.

\end{abstract}

\keywords{circumstellar matter, starspots, stars: activity, stars: low-mass, stars: winds, outflows}
\section{Introduction}

AU Mic is a nearby (9.8 pc; \citealt{gaia}), M1Ve star that is a member of the 23$\pm$3 Myr old $\beta$ Pic moving group \citep{mam14}. Since the discovery and early observations of the spatially resolved, edge-on debris disk associated with the system \citep{kalas04,liu04} that is gas poor \citep{aki2005}, subsequent high contrast imaging has probed the structure of disk in scattered light from projected stellocentric radii of 10 au \citep{wang2015} to 210 au \citep{glenn2014}, as well as distances in between \citep{met2005,krist2005,graham2007,fitz2007,boc15,boc18}. Along with longer wavelength mm observations, these data suggest the general architecture of AU Mic's debris system might be similar to the asteroid belt and Kuiper Belt in our own Solar System. AU Mic has a `birth ring' of material near 43 au that forms micron-sized particles from collisions between larger bodies (see e.g. \citealt{strubbe06,graham2007,mac13,matthews2015}), as well as an extended halo comprised of approximately micron-sized grains \citep{matthews2015}.  While it has been suggested that the inner (< 30 au) disk is devoid of small, micron-sized grains \citep{strubbe06}, recent observations suggest at least some small grains still exist here \citep{lomax18}. Early searches for planets in the system yielded null results \citep{hebb2007,met2005}; however, Plavchan+ (2019, submitted) presents the observational detection of the inner planetary population in the system. 

AU Mic is particularly notable in that it is the first spatially resolved debris disk where dynamical processes in the disk are being resolved on $<$ year timescales. \citet{boc15} analyzed multi-epoch imagery of the disk and discovered at least five spatially resolved features on the SE-side of the disk that clearly change location as a function of time, with measured projected tangential speeds of some features indicating they are moving on unbound, non-Keplerian trajectories.  Follow-up studies have identified additional moving features on both the SE- and NW-sides of the disk, and revealed that these features also exhibit vertical (perpendicular to the disk mid-plane) motion (e.g. \citealt{boc18} and Grady et al 2019, submitted). More recently, 
\citet{lomax18} noted a change in the color of the disk between 30-45au from being ``increasingly bluer with stellocentric distance'' \citep{krist2005,met2005,fitz2007} to a constant, smaller blue color, suggesting a reduction in the relative number of small grains at these stellocentric distances that could be causally correlated to the passage of fast-moving features seen in broad-band scattered light imagery.

The mechanism(s) driving the observed variability in AU Mic's debris disk remain hotly debated within the literature. \citet{sez17} suggest the moving dust features are either generated by resonance with a parent body that orbits at 8$\pm$2 au or at a recent large collision that generated a large population of smaller bodies, which are then dispersed by the stellar wind. \citet{cf17} proposed that the moving features in AU Mic's disk are caused by the interaction between the star's wind and repeated ``dust avalanche'' events. These avalanches are triggered in a zone marked by the intersection of AU Mic's primary debris ring and a proposed secondary ring of dust left behind by the catastrophic disruption of an object up to the size of the Kuiper Belt Object Varuna (radius ~450km; \citealt{lel2013}). Both \citet{cf17} and \citet{daley2019} discuss some of the potential challenges to this avalanche scenario. \citet{cf17} also predicts a vertical velocity component in moving features caused by the star's magnetized wind exerting a Lorentz force on the dust grains, potentially regulated by the stellar magnetic activity cycle.   These dynamical models assume significantly different wind properties: \citet{sez17} adopt a constant wind that induces a mass-loss rate that is 50x Solar, whereas \citet{cf17} adopt a variable wind that induces a mass-loss rate that is 500x-5000x Solar (see also \citealt{boc18}).  While \citet{au06} and \citet{sch15} detail why AU Mic's mass-loss rate is expected to be larger than that of the Sun; this mass-loss rate has not been observationally confirmed (e.g. \citealt{wood2005}). \citet{sez17} note that these dramatically different mass-loss rates have profound impacts on the grain blowout sizes that would be in operation.

In this paper, we present new white-light optical coronagraphic imagery of the inner region of AU Mic's disk obtained with the Hubble Space Telescope Imaging Spectrograph (HST/STIS) in 2018 using the BAR5 occulter. We describe the acquisition and reduction of these data in Section \ref{section:data}. We highlight the new morphologies of AU Mic's disk revealed by these data, including dramatic apparent loop-like structures seen in projection, in Section \ref{section:analysis}. We explore some of the potential mechanisms that could create these morphological features in the context of preliminary analysis of new constraints on the star's activity as enabled by NASA/TESS data in Section \ref{section:discussion}.

\section{HST BAR5 Coronagraphic Imagery} \label{section:data}

Coronagraphic observations of AU Mic and associated point spread function (PSF) template star HD 191849 were obtained with the HST/STIS BAR5 occulter, a 0$\farcs$15-wide rectangular bar, in GO-15219 (J. Wisniewski PI). The 50CORON imaging mode used provided an unfiltered spectral passband of pivot wavelength 0.575 $\mu$m with FWHM = 0.433 $\mu$m, an image scale of 50.77 mas pixel$^{-1}$, and spatial resolution of $\sim$72 mas. 

Observations of AU Mic were obtained in three sequential orbits on 26-27 July 2018 using a sub-array readout that sampled the inner 2$\farcs$5 (100 pixels) of the disk. The major axis of the disk was oriented approximately orthogonal to the long axis of the BAR5 occulter in the second orbit, and the telescope was rolled by $\pm$6.5$^{\circ}$ in the first and third orbits. Within each orbit, 20-21 images of 16.6 second duration were obtained at each of three cross-BAR5 dither positions offset by (-1/4, 0, +1/4) pixels orthogonally with respect to the midline of the occulter. HD 191849 was observed immediately thereafter using two sequential orbits, using the same cross-BAR5 dither technique, as a PSF reference. 22-26 images per dither location were obtained using 11.5 second per image integrations. We also obtained a single-orbit observation of AU Mic on 22 Sept 2018 with the disk aligned along the long axis of the BAR5 occulter, using the same dither and integration scheme described above. These disk-obscured images provide perfect color-matching to explore alternate methods to achieve better PSF subtraction. 

These data were reduced and calibrated using the same procedures and techniques as outlined in \citet{glenn2018}. Following bias, dark, and flat field correction using the temporally nearest calibration data, we located the position of the occulted star in each image using the "X marks the spot" diffraction spike fitting method of \citet{glenn2014}. Within each orbit, all images observed with the same dither position were then median combined and cleaned of cosmic-rays. Image co-registration, and later PSF subtractions, were done with sinc-apodized bi-cubic sub-pixel interpolation using the IDL-based, IDP3 software \citep{idp3}. The relative brightness and (x,y) position of the PSF star HD 191849 were treated, and iteratively adjusted, as free parameters to minimize the variance in difference image pixels not dominated by disk flux. 

We found that while chromatic residuals were fully mitigated with our disk-obscurred BAR5 observation of AU Mic on 22 Sept 2018, using this orbit as a PSF template led to substantial PSF subtraction residuals caused by differential wavefront errors induced by the non-contemporaneous (e.g. $\sim$2 month separation) of this observation from our 3 sequential orbits that resolved AU Mic's disk. We therefore adopt and use our sequentially contemporaneous observations of HD 191849 as the PSF reference for our final disk imagery. Following PSF-subtraction, our multi-roll images were re-oriented to a common orientation and median combined after masking known imaging artifacts.

\section{Analysis} \label{section:analysis}
\subsection{HST Imaging Results} \label{section:hstresults}
\begin{figure*}[t] 
	\begin{center}
		\includegraphics[width=\textwidth]{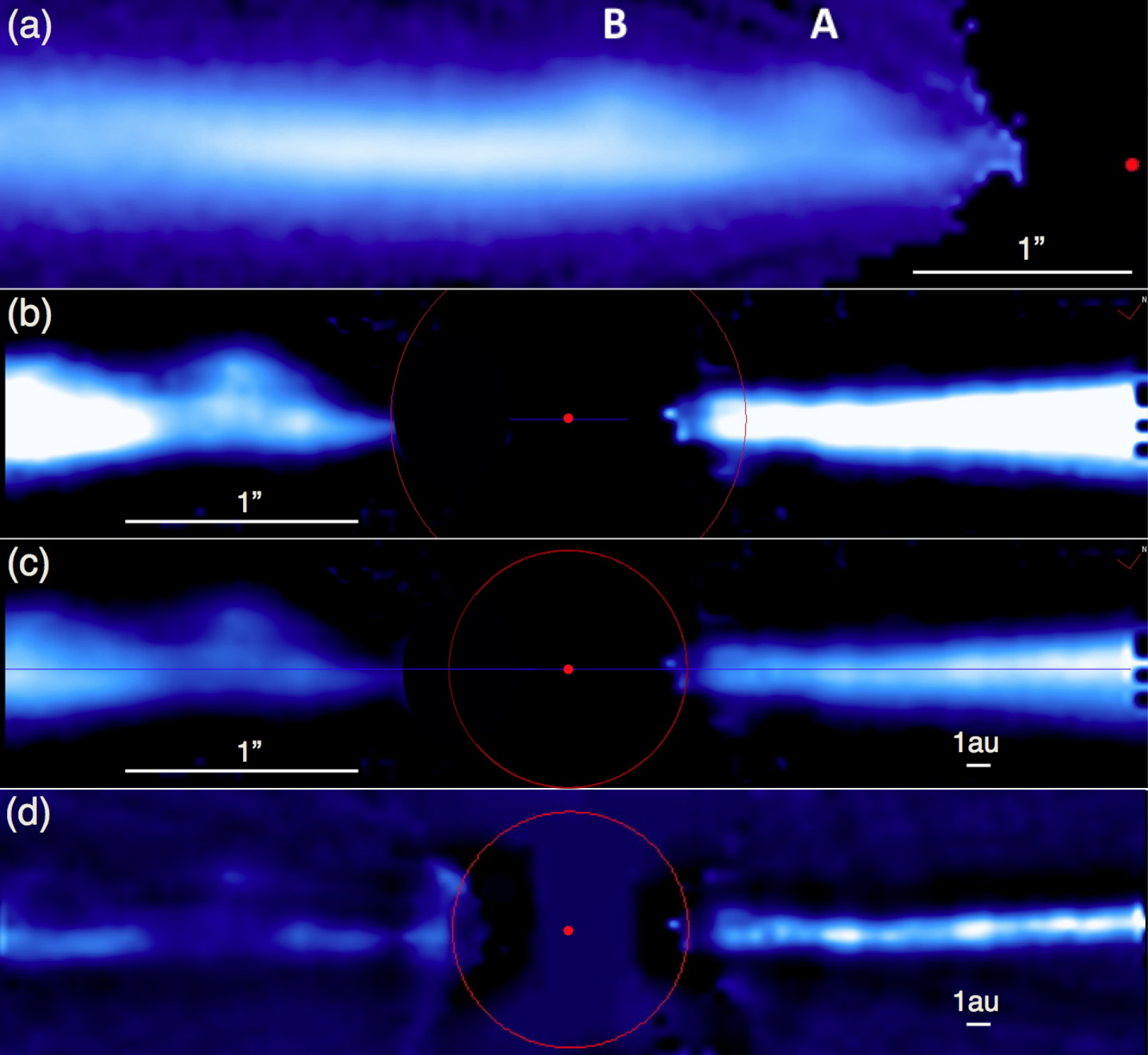}
	\end{center}
	\caption{The inner 5$\farcs$2 (50 au) region of the SE-side of the AU Mic disk from HST/STIS WedgeA0.6 imaging of the system in 2017 is shown in Panel (a), as adopted from Grady et al (2019, submitted). These data have been scaled by 1/r$^{2}$ and are displayed using a log scale optimized to emphasize the out-of-plane features A and B. Panels (b) and (c) depict HST/STIS BAR5 observations presented in this study, also scaled by 1/r$^{2}$. The FOV in both of these panels is 4$\farcs$93 x 1$\farcs$02, and the location of the central star is depicted by a red dot. Panels (b) and (c) adopted differently stretched log-based intensity scales to emphasize the dramatic ``loop-like'' morphology that feature A exhibits.  Applying a high-pass filter to the BAR5 data (panel d) enhances the visibility of disk sub-structure seen on au-scales.}
	\label{fig:bar5full}
\end{figure*}

\begin{figure}[t]
	\begin{center}
		\includegraphics[width=\columnwidth]{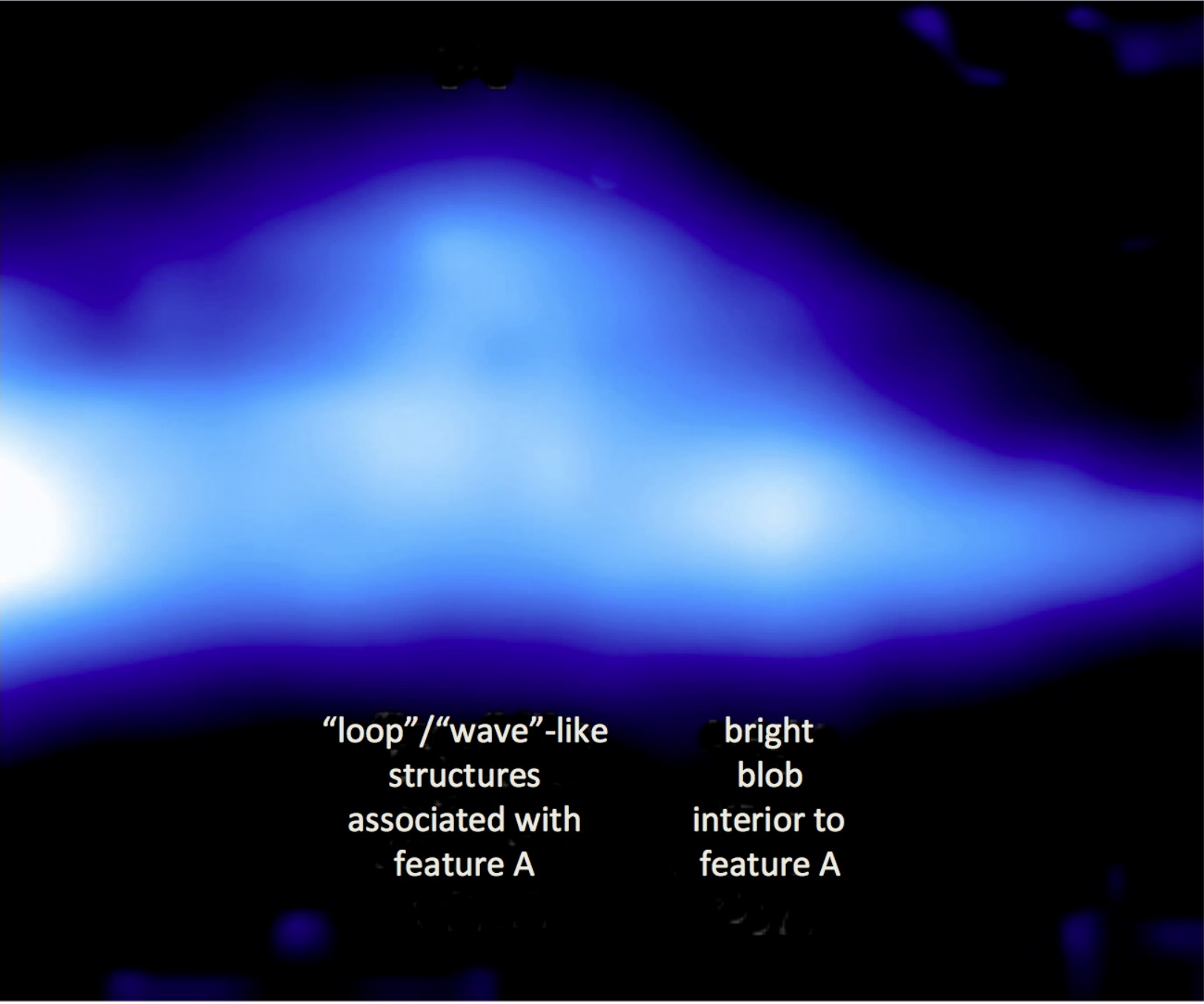}
	\end{center}
	\caption{The inner region of the SE-side of AU Mic's disk, spanning stellocentric separations of 0$\farcs$815 (right edge; 7.29 au) to 1$\farcs$881 (left edge; 18.28 au), is shown. The data are plotted on a log stretch after 1/r$^{2}$ scaling. Both the ``loop-like'' nature of feature A noted in Figure \ref{fig:bar5full} and a bright blob interior to this feature are apparent. As noted in Section \ref{section:hstresults}, the loop-like structure is located at a projected stellocentric separation of 14.2 au from AU Mic, and has a projected width of 1.5 au and rises 2.3 au above the disk midplane.}
	\label{fig:bar5loop}
\end{figure}

We display our resultant PSF-subtracted imagery of AU Mic taken with STIS' BAR5, after 1/r$^{2}$ intensity scaling, in panels b-d of Figure \ref{fig:bar5full}, along with 2017-epoch WedgeA0.6 imagery of the inner disk from Grady et al (2019, submitted) in panel a.  The superlative sub-pixel dithering of our BAR5 data provide an enhanced view of the morphology of feature A in the disk, compared to 2017-epoch WedgeA0.6 data. In particular, Figure \ref{fig:bar5full} (panels b-c) and Figure \ref{fig:bar5loop} reveals feature A has a distinctive ``loop-like'' appearance. It is not possible to determine whether the ``loop-like'' morphology of feature A remains coherent as the feature moves within the disk, due to the single-epoch nature of these high resolution data. As feature A is moving radially at $\sim$1.7 STIS pixels yr$^{-1}$ (i.e. $\sim$1.2 resolution elements yr$^{-1}$; Grady 2019, submitted), annual observations are commensurate with resolvable motions of this sub-structure. HST GO-15907 is pursuing these observations and could enable discrimination of the potential role of projection. Moreover, a larger portion of the disk will be read-out in these planned observations, allowing us to determine whether other known moving features in the disk exhibit loop-like morphologies similar to that found for feature A.  

To further explore new information encoded within these high-resolution BAR5 imagery, we applied a high-pass filter to these data (panel d, Figure \ref{fig:bar5full}). We fit two 1-dimensional Guassians to the loop-like structure to quantify its size and projected location.  These fits reveal the loop-like structure has a projected width of 1.5 au and rises to a projected height above the midplane of 2.3 au. The centroid of these fits also imply the loop-like structure resides at a projected stellocentric radial distance of 14.2 au from the host star.   

\subsection{TESS Photometry}
\begin{figure}[t]
	\begin{center}
		\includegraphics[width=\columnwidth]{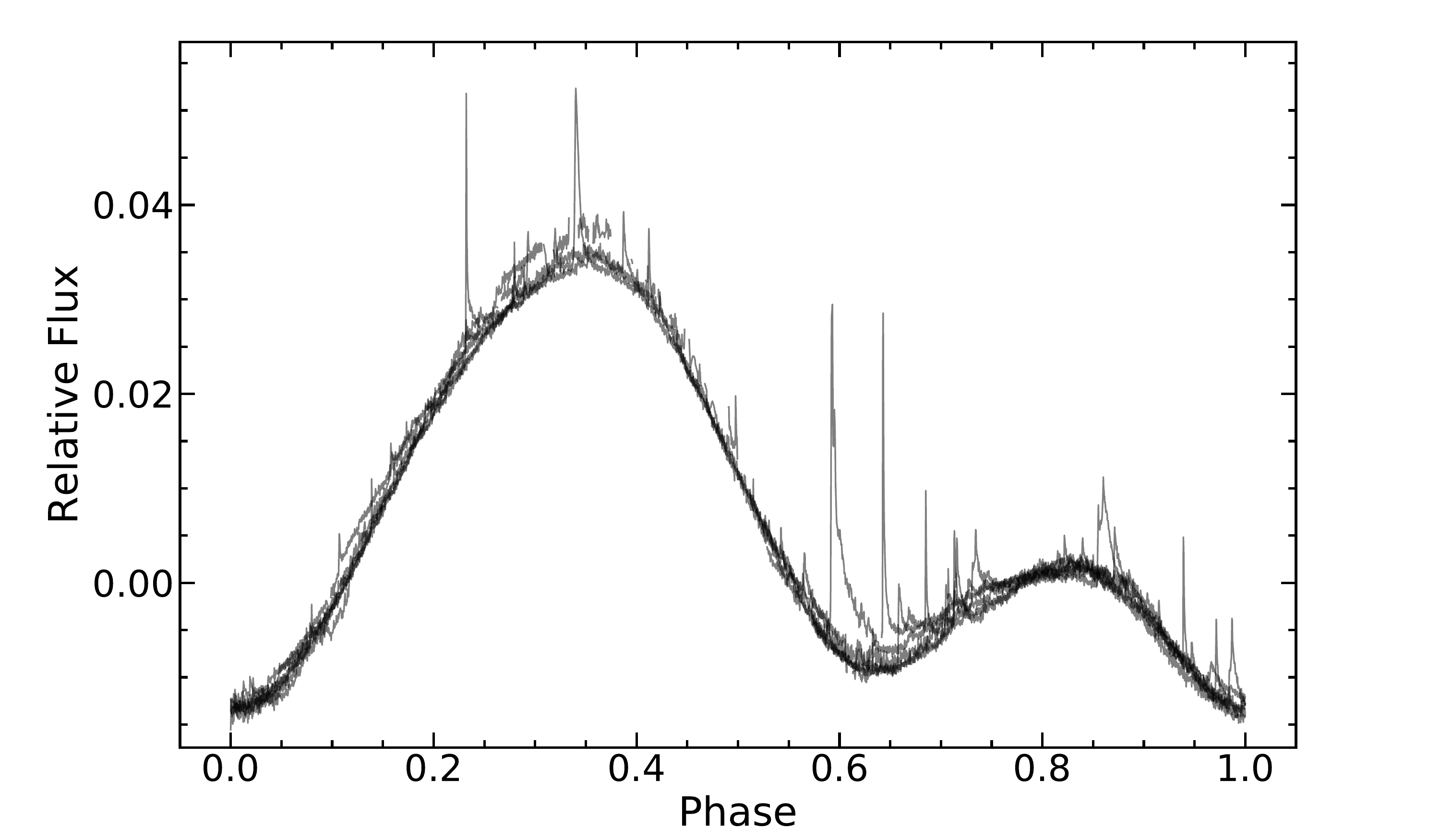}
	\end{center}
	\caption{The phase-folded light curve of TESS observations of AU Mic exhibit periodic modulation arising from star spots.}
	\label{fig:phase}
\end{figure}

AU Mic was observed at 2 minute cadence in Sector 1 (25 July 2018 - 22 August 2018) by the Transiting Exoplanet Survey Satellite (TESS) \citep{tess}. We find the data exhibit a 4.86 day periodicity arising from starspot-induced stellar activity, as seen in the phase-folded light curve shown in Figure \ref{fig:phase}. This best-fit periodicity is very similar to that derived from Lomb-Scargle fitting (4.875 days) \citep{lomb,scargle}. Two large, persistent spot complexes are clearly visible from the phase-folded light curve for $\sim$60\% and $\sim$40\% of 
the rotational period of the star respectively.

\subsection{Starspot Modeling} \label{spotsection}
Extracting robust information about the surface distribution of star spots and spot complexes from traditional spot modeling is often not possible as there exists a degeneracy between the stellar inclination and spot latitude \citep{w2013}. One can break this degeneracy in special cases, such as transiting planetary systems \citep{brett} or systems where the evolution of multiple spots can place weak constraints on their location \citep{davenport2015}. AU Mic is well known to have a spatially resolved, edge-on debris disk; hence, it represents another case where we have good confidence about the stellar inclination axis, assuming the disk's rotational axis is co-aligned. This 
assumption is supported by \citet{watson2011}, who report the inclination of the disk (90$^{\circ}$) to be the same as stellar rotational axis 
(90$^{\circ}{}^{+0}_{-20}$), and is consistent with the small level of
mutual inclination between the disk and stellar rotational axis 
($\mid \Delta i \mid = 1^{\circ}{}^{+7}_{-1}$) found 
by \citet{greaves2014}. Even in debris disk systems that have resolved disks and directly imaged planets with evidence of mutual misalignment, such as beta Pic b, the amplitude of this misalignment is 
small ($\sim$3 $^{\circ}$).

\citet{w2013} showed that the fraction of time that spots are visible in a system whose stellar inclination is seen edge-on, i.e. i = 90$^{\circ}$, is a nearly uniform $\sim$55\% of its rotational period, that rises to 60\% near polar latitudes.  Figure \ref{fig:phase} demonstrates that the smaller amplitude spot complex in AU Mic's phase-folded light curve persists for $\sim$40\% of the star's rotational period, which is not expected for standard edge-on systems \citep{w2013}.  Resolving this discrepancy demands relaxing the fundamental assumption of \citet{w2013} that the rotational and magnetic field axes are co-aligned. 

We utilize the starspot modeling software \texttt{STSP} developed by L. Hebb (2019, in preparation), as described in detail within \citet{davenport2015}, to model the TESS data. This software generates synthetic light curves for a star having a pre-defined number of static spots (or spot complexes), and computes spot properties (latitude, longitude, radius) from a $\chi^{2}$ comparison between computed synthetic fluxes and observed fluxes using a Markov Chain Monte Carlo routine based on \citet{mcmc}. We used 50 parameter space walkers, with 20,000 steps in the MCMC chain.

We further explored the possibility of a mis-aligned B field in the system via a proxy, namely by varying the star's ``rotation axis'' using a grid of MCMC runs from 90$^\circ$ (i.e. edge-on) to 30$^\circ$ in steps of $15^\circ$.  Finding evidence of a preferred non-edge-on orientation via this modeling (given the evidence we have that the stellar rotational axis is co-aligned with the edge-on disk) serves as a proxy for the potential B field mis-alignment. Full 20,000 step MCMC explorations were independently run for each inclination, and the best (lowest $\chi^2$) solution from the converged portion of the chains was considered. This grid found the rotation axis is weakly constrained from the starspot data alone, with the best model of $i = 75^\circ$ being only slightly preferred ($\Delta \chi^2 \sim 1$) over the edge-on solution.
The \texttt{STSP} models more strongly rule out highly inclined scenarios of $45^\circ$ and $30^\circ$. Thus, since we know the system inclination is likely edge-on from our spatially resolved imagery, the \texttt{STSP} results support the suggestion that the effective B field of AU Mic is mis-alignment with its rotational axis. Further quantifying the detailed topology of AU Mic's B field via Zeeman-Doppler Imaging is strongly encouraged.

\begin{figure*}[t] 
	\begin{center}
		\includegraphics[width=6in]{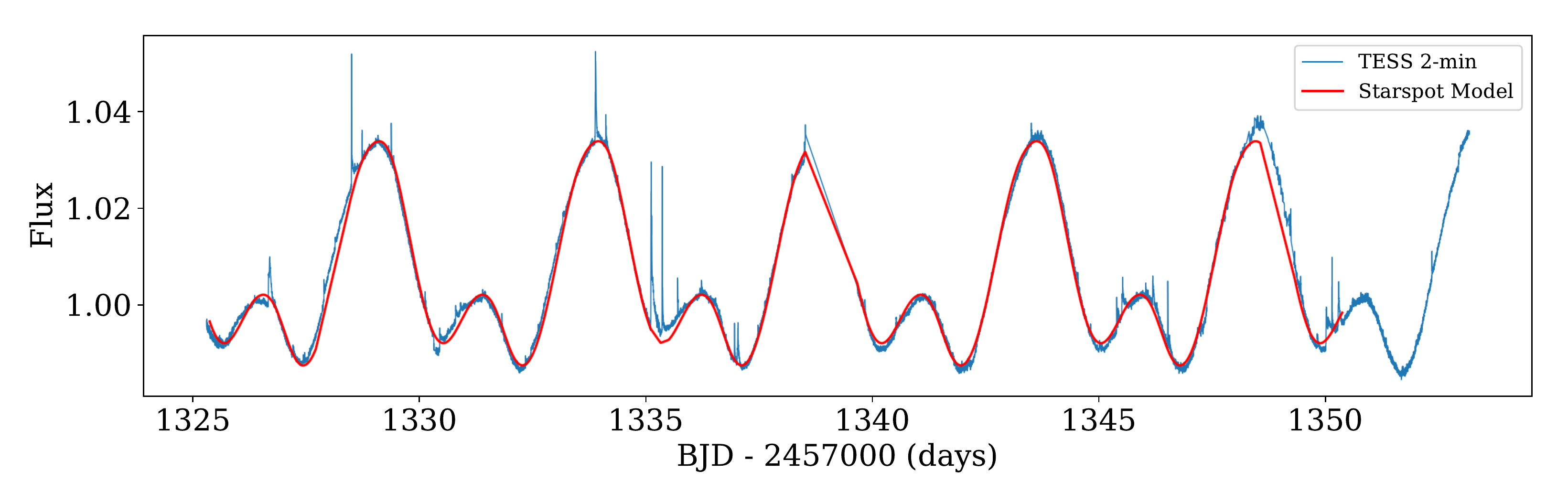}\\
		\includegraphics[width=3in]{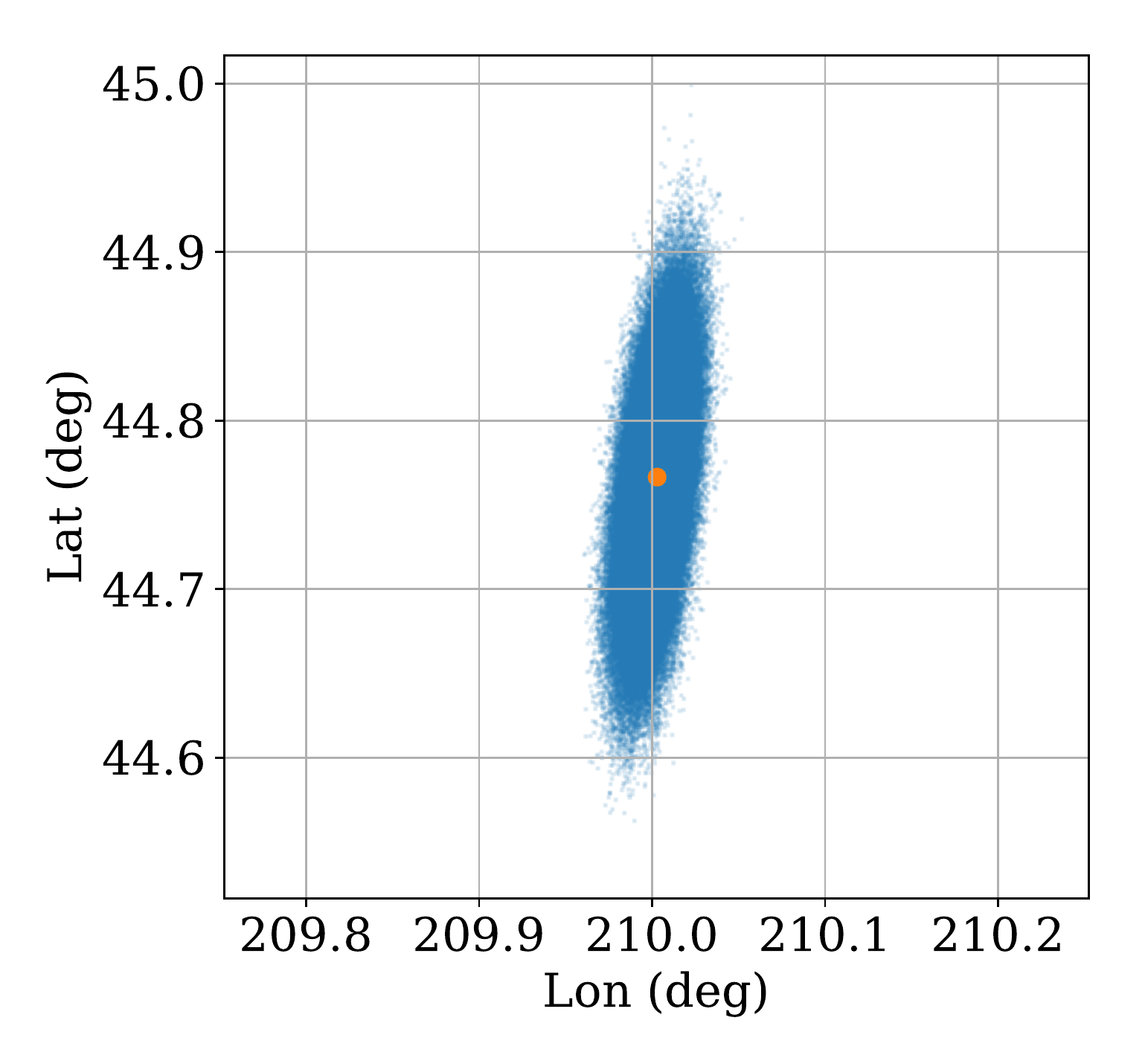}
		\includegraphics[width=3in]{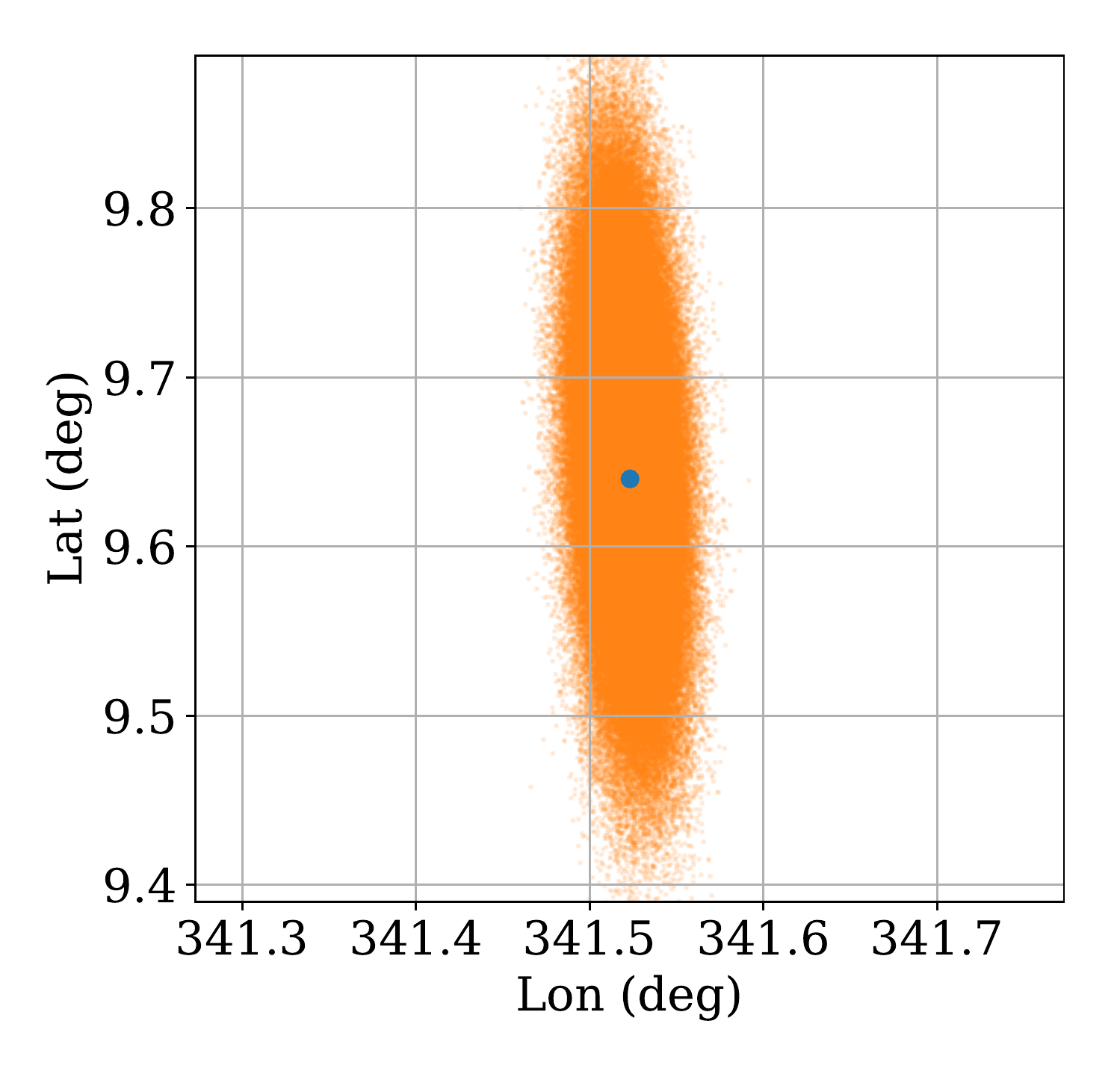}
	\end{center}
	\caption{Top: TESS 2-minute light curve for AU Mic (blue) with sinusoidal starspot modulations and flares. Our best fitting (lowest $\chi^2$) stationary 2 starspot model from {\tt STSP} is overlaid (red).
	Bottom: Latitude and longitude locations of the two starspots from the converged portions of our affine-invarient MCMC chains, which trace the posterior probability distribution. The lowest $\chi^2$ point is highlighted for each spot, which correlates to the model light curve above.
	}
	\label{fig:spotmodel}
\end{figure*}

The resultant best fit synthetic spot modeling light curve is shown in Figure \ref{fig:spotmodel} (red), along with the underlying TESS data (blue). The most likely relative latitude/longitude distribution of the two spot complexes in AU Mic are depicted in the bottom panels of Figure \ref{fig:spotmodel}. We find the separation of the spots in latitude is significant, with one spot near the equator ($\sim$9$^\circ$) and the second spot at higher latitutde ($\sim$44$^\circ$). The spots are separated by $\sim131^\circ$ in longitude.

\section{Discussion} \label{section:discussion}
\subsection{Origin of AU Mic's Loop-like Disk Structure}
Multi-epoch ground-based near-IR coronagraphic imagery of AU Mic's disk \citep{boc18}, confirmed by higher photometric fidelity multi-epoch space-based optical coronagraphic imagery (Grady et al 2019, submitted) has revealed evidence that the projected distribution of disk material evolves on both the projected vertical and radial directions. The detailed vertical evolution of these structures has been hard to quantify, given the sparse sampling of available high fidelity observations. The vertical extension and evolution of these features necessarily requires the presence of non-radial forces. \citet{cf17} suggested, for example, that the star's magnetized wind could exhibit a Lorentz force on grains and produce the vertical unduluations observed in spatially resolved imagery if the polarity of the stellar magnetic field reverses periodically. The striking projected structure that our new HST imagery has revealed for feature A (Figures \ref{fig:bar5full} - \ref{fig:bar5loop}), i.e. a ``loop-like'' morphology, is suggestive that material is actively being lifted both above the midplane and back down towards the midplane on short time-scales (i.e. significantly shorter than what could be caused by a change in the stellar magnetic field polarity). One potential way to produce such morphologies on this time-scale is if the disk is sculpted by a time-variable stellar wind. As discussed below, we suggest that a mis-alignment between the stellar B field and disk rotational axis (that is co-aligned with the stellar rotational axis) and/or a mis-alignment between the stellar rotational axis (co-aligned with the B field axis) and disk rotational axis could lead to the disk seeing different amounts of polar vs equatorial wind over time, producing morphological structure within the disk.

\subsection{Evidence of Mis-aligned B Field and Potential Ramifications} \label{borient}

We have shown in Section \ref{spotsection} that AU Mic's edge-on debris disk allows a plausible way to break the spot latitude vs. stellar inclination degeneracy and place constraints on the latitude of spots in the system (Figure \ref{fig:spotmodel}). \citet{w2013} aptly demonstrated that the fraction of time that spots are visible in a system whose stellar inclination is seen edge-on, i.e. i = 90$^{\circ}$, is a nearly uniform $\sim$55\% of its rotational period, that rises to 60\% near polar latitudes. AU Mic's phase-folded light curve (Figure \ref{fig:phase}) clearly violates this expectation, as its weaker amplitude spot is visible for $\sim$40\% of its rotational phase. Since we know AU Mic's stellar rotational axis with high confidence from its resolved disk (and studies that indicate minimal mis-alignment between the disk and stellar rotational axis, see e.g. \citealt{watson2011,greaves2014}), this implies that it must have an offset magnetic field axis. Indeed, our spot modeling in Section \ref{spotsection} supported this offset magnetic field interpretation. Offset magnetic field axes have been proposed in other dM systems like GJ 1243 \citep{davenport2015}, and other astrophysical systems like magnetically confined disks surrounding massive B-type stars \citep{rrm1,rrm2}.

We consider the potential implications of AU Mic having a magnetic field axis that is offset from its rotational axis, and in particular the implications that this could have on its extended debris disk. \citet{vidotto2014} found that non-axisymmetric surface magnetic fields can lead to more asymmetric mass fluxes.  A common feature of the two leading dynamical models to explain the super-Keplerian motion of moving features in AU Mic's disk is that both invoke the system's stellar wind to disperse dust grains \citep{sez17,cf17}.  We speculate that small grains in AU Mic's disk could be sculpted by a time-dependent wind, that could not only be enhanced by potentially large coronal mass ejections \citep{boc15} but also be influenced by the system having an offset B field axis. A potential analogy in the Sun are the interactions between the fast Solar wind, that exits through coronal holes, and the slower Solar wind, which creates stream interaction regions that co-rotate with the Sun \citep{sun}. The effects of these Solar interaction regions have been observed by the Voyager 1 and 2 and Pioneer 10 missions, including at stellocentric separations from the Sun ($\sim$16 au) similar to the current projected location of feature A in AU Mic \citep{voyager,voyager2,voyager3,sun}. These time-dependent interaction regions create large-scale spiral density features that affect both the radial and vertical distribution of material in the Heliosphere.  If analogous interaction regions exist around AU Mic, this could serve as one mechanism that could contribute to shaping the projected radial and vertical evolution of dust grains in AU Mic's disk, as diagnosed by spatially resolved imagery. Although beyond the scope of this Letter, detailed dynamical modeling of such co-rotating interaction regions in the AU Mic system should be pursued. It could also be interesting to explore the magnetic field structure of other M-type debris disks that exhibit potential spiral-like structures, such as TWA 7 \citep{olofsson}, to help assess whether they exhibit similarities to the AU Mic system.

\subsection{Alternate Forms of Mis-alignment and Potential Ramifications}
Our spot modeling (Section \ref{spotsection}) and the associated interpretation of these results (Section \ref{borient}) utilized previous research and observational properties of the system that indicated co-alignment of AU Mic's stellar and disk rotational axes. We remark that if we relax this co-alignment prior, the observed phase-dependence of AU Mic's spot coverage (two spot complexes visible for $\sim$60\% and $\sim$40\% of the rotational period respectively) would demand mis-alignment between the edge-on disk and the stellar rotational axis.  In such a scenario, our available data would not place constraints on the relative orientation of the B field.  However, this mis-alignment would expose the disk to seeing different amounts of the equatorial versus more polar wind with each rotation of the star.  We thus suggest that small grains in AU Mic's disk could be similarly sculpted by such a time-dependent wind.

\subsection{Using Edge-on Disks to Break Starspot Modeling Degeneracies} 
The advent of high precision, high cadence, long duration photometric datasets from space-based missions like Kepler and TESS provide a unique opportunity to identify and characterize stellar activity arising from spot modulation across a large sample of low mass stars.  We remark that our work with AU Mic reveals another unique way to break the degeneracy between stellar inclination and spot latitudes. Much like the special case of transiting planetary systems, whereby the presence of a transit allows one to deduce the stellar inclination angle with high precision \citep{brett}, the subset of circumstellar disk systems that offer detailed inclination angle information (e.g. as measured from spatially resolved imaging for both edge-on, face-on, and intermediate inclination systems, or as inferred from photometric behavior like so-called ``dipper'' systems, which could suggest a likely near edge-on inclination \citep{cody2014}) could provide another way to break key degeneracies in spot modeling.  In particular, leveraging system inclinations from disk properties could enable one to map the detailed distribution of starspots and the prevalence of offset magnetic field axes from a large statistical sample of stars observed by Kepler/TESS.

\acknowledgments
We thank our referee for providing feedback that helped to improve both the clarity and content of this manuscript. This work was supported by a grant from STScI for GO-15219. We thank Suzanne Hawley, Jamie Lomax, Peter Plavchan, and Ben Tofflemire for helpful discussions of this work.


\pagebreak
\bibliographystyle{yahapj}

\bibliography{main}

\begin{thebibliography}{}
\providecommand\natexlab[1]{#1}
\providecommand\JournalTitle[1]{#1}

\bibitem[{{Augereau} \& {Beust}(2006)}]{au06}
{Augereau}, J.-C., \& {Beust}, H. 2006,
  \href{http://dx.doi.org/10.1051/0004-6361:20054250}{\JournalTitle{\aap}, 455,
  987}

\bibitem[{{Boccaletti} {et~al.}(2015){Boccaletti}, {Thalmann}, {Lagrange},
  {Janson}, {Augereau}, {Schneider}, {Milli}, {Grady}, {Debes}, {Langlois},
  {Mouillet}, {Henning}, {Dominik}, {Maire}, {Beuzit}, {Carson}, {Dohlen},
  {Engler}, {Feldt}, {Fusco}, {Ginski}, {Girard}, {Hines}, {Kasper}, {Mawet},
  {M{\'e}nard}, {Meyer}, {Moutou}, {Olofsson}, {Rodigas}, {Sauvage},
  {Schlieder}, {Schmid}, {Turatto}, {Udry}, {Vakili}, {Vigan}, {Wahhaj}, \&
  {Wisniewski}}]{boc15}
{Boccaletti}, A., {Thalmann}, C., {Lagrange}, A.-M., {et~al.} 2015,
  \href{http://dx.doi.org/10.1038/nature15705}{\JournalTitle{\nat}, 526, 230}

\bibitem[{{Boccaletti} {et~al.}(2018){Boccaletti}, {Sezestre}, {Lagrange},
  {Th{\'e}bault}, {Gratton}, {Langlois}, {Thalmann}, {Janson}, {Delorme},
  {Augereau}, {Schneider}, {Milli}, {Grady}, {Debes}, {Kral}, {Olofsson},
  {Carson}, {Maire}, {Henning}, {Wisniewski}, {Schlieder}, {Dominik},
  {Desidera}, {Ginski}, {Hines}, {M{\'e}nard}, {Mouillet}, {Pawellek}, {Vigan},
  {Lagadec}, {Avenhaus}, {Beuzit}, {Biller}, {Bonavita}, {Bonnefoy},
  {Brandner}, {Cantalloube}, {Chauvin}, {Cheetham}, {Cudel}, {Gry}, {Daemgen},
  {Feldt}, {Galicher}, {Girard}, {Hagelberg}, {Janin-Potiron}, {Kasper}, {Le
  Coroller}, {Mesa}, {Peretti}, {Perrot}, {Samland}, {Sissa}, {Wildi}, {Zurlo},
  {Rochat}, {Stadler}, {Gluck}, {Orign{\'e}}, {Llored}, {Baudoz}, {Rousset},
  {Martinez}, \& {Rigal}}]{boc18}
{Boccaletti}, A., {Sezestre}, E., {Lagrange}, A.-M., {et~al.} 2018,
  \href{http://dx.doi.org/10.1051/0004-6361/201732462}{\JournalTitle{\aap},
  614, A52}

\bibitem[{{Burlaga}(1988)}]{voyager2}
{Burlaga}, L.~F. 1988,
  \href{http://dx.doi.org/10.1029/JA093iA05p04103}{\JournalTitle{\jgr}, 93,
  4103}

\bibitem[{{Burlaga} {et~al.}(1984){Burlaga}, {McDonald}, {Ness}, {Schwenn},
  {Lazarus}, \& {Mariani}}]{voyager}
{Burlaga}, L.~F., {McDonald}, F.~B., {Ness}, N.~F., {et~al.} 1984,
  \href{http://dx.doi.org/10.1029/JA089iA08p06579}{\JournalTitle{\jgr}, 89,
  6579}

\bibitem[{{Chiang} \& {Fung}(2017)}]{cf17}
{Chiang}, E., \& {Fung}, J. 2017,
  \href{http://dx.doi.org/10.3847/1538-4357/aa89e6}{\JournalTitle{\apj}, 848,
  4}

\bibitem[{{Cody} {et~al.}(2014){Cody}, {Stauffer}, {Baglin}, {Micela},
  {Rebull}, {Flaccomio}, {Morales-Calder{\'o}n}, {Aigrain}, {Bouvier}, \&
  {Hillenbrand}}]{cody2014}
{Cody}, A.~M., {Stauffer}, J., {Baglin}, A., {et~al.} 2014,
  \href{http://dx.doi.org/10.1088/0004-6256/147/4/82}{\JournalTitle{\aj}, 147,
  82}

\bibitem[{{Daley} {et~al.}(2019){Daley}, {Hughes}, {Carter}, {Flaherty},
  {Lambros}, {Pan}, {Schlichting}, {Chiang}, {Wyatt}, \& {Wilner}}]{daley2019}
{Daley}, C., {Hughes}, A.~M., {Carter}, E.~S., {et~al.} 2019,
  \href{http://dx.doi.org/10.3847/1538-4357/ab1074}{\JournalTitle{\apj}, 875,
  87}

\bibitem[{{Davenport} {et~al.}(2015){Davenport}, {Hebb}, \&
  {Hawley}}]{davenport2015}
{Davenport}, J.~R.~A., {Hebb}, L., \& {Hawley}, S.~L. 2015,
  \href{http://dx.doi.org/10.1088/0004-637X/806/2/212}{\JournalTitle{\apj},
  806, 212}

\bibitem[{{Fitzgerald} {et~al.}(2007){Fitzgerald}, {Kalas}, {Duch{\^e}ne},
  {Pinte}, \& {Graham}}]{fitz2007}
{Fitzgerald}, M.~P., {Kalas}, P.~G., {Duch{\^e}ne}, G., {Pinte}, C., \&
  {Graham}, J.~R. 2007,
  \href{http://dx.doi.org/10.1086/521344}{\JournalTitle{\apj}, 670, 536}

\bibitem[{{Foreman-Mackey} {et~al.}(2013){Foreman-Mackey}, {Hogg}, {Lang}, \&
  {Goodman}}]{mcmc}
{Foreman-Mackey}, D., {Hogg}, D.~W., {Lang}, D., \& {Goodman}, J. 2013,
  \href{http://dx.doi.org/10.1086/670067}{\JournalTitle{\pasp}, 125, 306}

\bibitem[{{Gaia Collaboration} {et~al.}(2018){Gaia Collaboration}, {Brown},
  {Vallenari}, {Prusti}, {de Bruijne}, {Babusiaux}, {Bailer-Jones}, {Biermann},
  {Evans}, {Eyer}, \& et~al.}]{gaia}
{Gaia Collaboration}, {Brown}, A.~G.~A., {Vallenari}, A., {et~al.} 2018,
  \href{http://dx.doi.org/10.1051/0004-6361/201833051}{\JournalTitle{\aap},
  616, A1}

\bibitem[{{Gazis} {et~al.}(1999){Gazis}, {McDonald}, {Burger}, {Chalov},
  {Decker}, {Dwyer}, {Intriligator}, {Jokipii}, {Lazarus}, {Mason}, {Pizzo},
  {Potgieter}, {Richardson}, \& {Lanzerotti}}]{voyager3}
{Gazis}, P.~R., {McDonald}, F.~B., {Burger}, R.~A., {et~al.} 1999,
  \href{http://dx.doi.org/10.1023/A:1005270027347}{\JournalTitle{\ssr}, 89,
  269}

\bibitem[{{Graham} {et~al.}(2007){Graham}, {Kalas}, \& {Matthews}}]{graham2007}
{Graham}, J.~R., {Kalas}, P.~G., \& {Matthews}, B.~C. 2007,
  \href{http://dx.doi.org/10.1086/509318}{\JournalTitle{\apj}, 654, 595}

\bibitem[{{Greaves} {et~al.}(2014){Greaves}, {Kennedy}, {Thureau}, {Eiroa},
  {Marshall}, {Maldonado}, {Matthews}, {Olofsson}, {Barlow},
  {Moro-Mart{\'\i}n}, {Sibthorpe}, {Absil}, {Ardila}, {Booth},
  {Broekhoven-Fiene}, {Brown}, {Cameron}, {del Burgo}, {Di Francesco},
  {Eisl{\"o}ffel}, {Duch{\^e}ne}, {Ertel}, {Holland}, {Horner}, {Kalas},
  {Kavelaars}, {Lestrade}, {Vican}, {Wilner}, {Wolf}, \& {Wyatt}}]{greaves2014}
{Greaves}, J.~S., {Kennedy}, G.~M., {Thureau}, N., {et~al.} 2014,
  \href{http://dx.doi.org/10.1093/mnrasl/slt153}{\JournalTitle{\mnras}, 438,
  L31}

\bibitem[{{Hebb} {et~al.}(2007){Hebb}, {Petro}, {Ford}, {Ardila}, {Toledo},
  {Minniti}, {Golimowski}, \& {Clampin}}]{hebb2007}
{Hebb}, L., {Petro}, L., {Ford}, H.~C., {et~al.} 2007,
  \href{http://dx.doi.org/10.1111/j.1365-2966.2007.11904.x}{\JournalTitle{\mnras},
  379, 63}

\bibitem[{{Kalas} {et~al.}(2004){Kalas}, {Liu}, \& {Matthews}}]{kalas04}
{Kalas}, P., {Liu}, M.~C., \& {Matthews}, B.~C. 2004,
  \href{http://dx.doi.org/10.1126/science.1093420}{\JournalTitle{Science}, 303,
  1990}

\bibitem[{{Krist} {et~al.}(2005){Krist}, {Ardila}, {Golimowski}, {Clampin},
  {Ford}, {Illingworth}, {Hartig}, {Bartko}, {Ben{\'{\i}}tez}, {Blakeslee},
  {Bouwens}, {Bradley}, {Broadhurst}, {Brown}, {Burrows}, {Cheng}, {Cross},
  {Demarco}, {Feldman}, {Franx}, {Goto}, {Gronwall}, {Holden}, {Homeier},
  {Infante}, {Kimble}, {Lesser}, {Martel}, {Mei}, {Menanteau}, {Meurer},
  {Miley}, {Motta}, {Postman}, {Rosati}, {Sirianni}, {Sparks}, {Tran},
  {Tsvetanov}, {White}, \& {Zheng}}]{krist2005}
{Krist}, J.~E., {Ardila}, D.~R., {Golimowski}, D.~A., {et~al.} 2005,
  \href{http://dx.doi.org/10.1086/426755}{\JournalTitle{\aj}, 129, 1008}

\bibitem[{{Lellouch} {et~al.}(2013){Lellouch}, {Santos-Sanz}, {Lacerda},
  {Mommert}, {Duffard}, {Ortiz}, {M{\"u}ller}, {Fornasier}, {Stansberry},
  {Kiss}, {Vilenius}, {Mueller}, {Peixinho}, {Moreno}, {Groussin}, {Delsanti},
  \& {Harris}}]{lel2013}
{Lellouch}, E., {Santos-Sanz}, P., {Lacerda}, P., {et~al.} 2013,
  \href{http://dx.doi.org/10.1051/0004-6361/201322047}{\JournalTitle{\aap},
  557, A60}

\bibitem[{{Liu}(2004)}]{liu04}
{Liu}, M.~C. 2004,
  \href{http://dx.doi.org/10.1126/science.1102929}{\JournalTitle{Science}, 305,
  1442}

\bibitem[{{Lomax} {et~al.}(2018){Lomax}, {Wisniewski}, {Roberge}, {Donaldson},
  {Debes}, {Malumuth}, \& {Weinberger}}]{lomax18}
{Lomax}, J.~R., {Wisniewski}, J.~P., {Roberge}, A., {et~al.} 2018,
  \href{http://dx.doi.org/10.3847/1538-3881/aaa1a7}{\JournalTitle{\aj}, 155,
  62}

\bibitem[{{Lomb}(1976)}]{lomb}
{Lomb}, N.~R. 1976,
  \href{http://dx.doi.org/10.1007/BF00648343}{\JournalTitle{\apss}, 39, 447}

\bibitem[{{MacGregor} {et~al.}(2013){MacGregor}, {Wilner}, {Rosenfeld},
  {Andrews}, {Matthews}, {Hughes}, {Booth}, {Chiang}, {Graham}, {Kalas},
  {Kennedy}, \& {Sibthorpe}}]{mac13}
{MacGregor}, M.~A., {Wilner}, D.~J., {Rosenfeld}, K.~A., {et~al.} 2013,
  \href{http://dx.doi.org/10.1088/2041-8205/762/2/L21}{\JournalTitle{\apjl},
  762, L21}

\bibitem[{{Mamajek} \& {Bell}(2014)}]{mam14}
{Mamajek}, E.~E., \& {Bell}, C.~P.~M. 2014,
  \href{http://dx.doi.org/10.1093/mnras/stu1894}{\JournalTitle{\mnras}, 445,
  2169}

\bibitem[{{Matthews} {et~al.}(2015){Matthews}, {Kennedy}, {Sibthorpe},
  {Holland}, {Booth}, {Kalas}, {MacGregor}, {Wilner}, {Vandenbussche},
  {Olofsson}, {Blommaert}, {Brandeker}, {Dent}, {de Vries}, {Di Francesco},
  {Fridlund}, {Graham}, {Greaves}, {Heras}, {Hogerheijde}, {Ivison}, {Pantin},
  \& {Pilbratt}}]{matthews2015}
{Matthews}, B.~C., {Kennedy}, G., {Sibthorpe}, B., {et~al.} 2015,
  \href{http://dx.doi.org/10.1088/0004-637X/811/2/100}{\JournalTitle{\apj},
  811, 100}

\bibitem[{{Metchev} {et~al.}(2005){Metchev}, {Eisner}, {Hillenbrand}, \&
  {Wolf}}]{met2005}
{Metchev}, S.~A., {Eisner}, J.~A., {Hillenbrand}, L.~A., \& {Wolf}, S. 2005,
  \href{http://dx.doi.org/10.1086/427869}{\JournalTitle{\apj}, 622, 451}

\bibitem[{{Morris} {et~al.}(2017){Morris}, {Hebb}, {Davenport}, {Rohn}, \&
  {Hawley}}]{brett}
{Morris}, B.~M., {Hebb}, L., {Davenport}, J.~R.~A., {Rohn}, G., \& {Hawley},
  S.~L. 2017,
  \href{http://dx.doi.org/10.3847/1538-4357/aa8555}{\JournalTitle{\apj}, 846,
  99}

\bibitem[{{Olofsson} {et~al.}(2018){Olofsson}, {van Holstein}, {Boccaletti},
  {Janson}, {Th{\'e}bault}, {Gratton}, {Lazzoni}, {Kral}, {Bayo}, \&
  {Canovas}}]{olofsson}
{Olofsson}, J., {van Holstein}, R.~G., {Boccaletti}, A., {et~al.} 2018,
  \href{http://dx.doi.org/10.1051/0004-6361/201832583}{\JournalTitle{\aap},
  617, A109}

\bibitem[{{Richardson}(2018)}]{sun}
{Richardson}, I.~G. 2018,
  \href{http://dx.doi.org/10.1007/s41116-017-0011-z}{\JournalTitle{Living
  Reviews in Solar Physics}, 15, 1}

\bibitem[{{Ricker} {et~al.}(2015){Ricker}, {Winn}, {Vanderspek}, {Latham},
  {Bakos}, {Bean}, {Berta-Thompson}, {Brown}, {Buchhave}, {Butler}, {Butler},
  {Chaplin}, {Charbonneau}, {Christensen-Dalsgaard}, {Clampin}, {Deming},
  {Doty}, {De Lee}, {Dressing}, {Dunham}, {Endl}, {Fressin}, {Ge}, {Henning},
  {Holman}, {Howard}, {Ida}, {Jenkins}, {Jernigan}, {Johnson}, {Kaltenegger},
  {Kawai}, {Kjeldsen}, {Laughlin}, {Levine}, {Lin}, {Lissauer}, {MacQueen},
  {Marcy}, {McCullough}, {Morton}, {Narita}, {Paegert}, {Palle}, {Pepe},
  {Pepper}, {Quirrenbach}, {Rinehart}, {Sasselov}, {Sato}, {Seager},
  {Sozzetti}, {Stassun}, {Sullivan}, {Szentgyorgyi}, {Torres}, {Udry}, \&
  {Villasenor}}]{tess}
{Ricker}, G.~R., {Winn}, J.~N., {Vanderspek}, R., {et~al.} 2015,
  \href{http://dx.doi.org/10.1117/1.JATIS.1.1.014003}{\JournalTitle{Journal of
  Astronomical Telescopes, Instruments, and Systems}, 1, 014003}

\bibitem[{{Roberge} {et~al.}(2005){Roberge}, {Weinberger}, {Redfield}, \&
  {Feldman}}]{aki2005}
{Roberge}, A., {Weinberger}, A.~J., {Redfield}, S., \& {Feldman}, P.~D. 2005,
  \href{http://dx.doi.org/10.1086/431899}{\JournalTitle{\apjl}, 626, L105}

\bibitem[{{Scargle}(1982)}]{scargle}
{Scargle}, J.~D. 1982,
  \href{http://dx.doi.org/10.1086/160554}{\JournalTitle{\apj}, 263, 835}

\bibitem[{{Schneider} {et~al.}(2014){Schneider}, {Grady}, {Hines}, {Stark},
  {Debes}, {Carson}, {Kuchner}, {Perrin}, {Weinberger}, {Wisniewski},
  {Silverstone}, {Jang-Condell}, {Henning}, {Woodgate}, {Serabyn},
  {Moro-Martin}, {Tamura}, {Hinz}, \& {Rodigas}}]{glenn2014}
{Schneider}, G., {Grady}, C.~A., {Hines}, D.~C., {et~al.} 2014,
  \href{http://dx.doi.org/10.1088/0004-6256/148/4/59}{\JournalTitle{\aj}, 148,
  59}

\bibitem[{{Schneider} {et~al.}(2018){Schneider}, {Debes}, {Grady},
  {G{\'a}sp{\'a}r}, {Henning}, {Hines}, {Kuchner}, {Perrin}, \&
  {Wisniewski}}]{glenn2018}
{Schneider}, G., {Debes}, J.~H., {Grady}, C.~A., {et~al.} 2018,
  \href{http://dx.doi.org/10.3847/1538-3881/aaa3f3}{\JournalTitle{\aj}, 155,
  77}

\bibitem[{{Sch{\"u}ppler} {et~al.}(2015){Sch{\"u}ppler}, {L{\"o}hne}, {Krivov},
  {Ertel}, {Marshall}, {Wolf}, {Wyatt}, {Augereau}, \& {Metchev}}]{sch15}
{Sch{\"u}ppler}, C., {L{\"o}hne}, T., {Krivov}, A.~V., {et~al.} 2015,
  \href{http://dx.doi.org/10.1051/0004-6361/201525664}{\JournalTitle{\aap},
  581, A97}

\bibitem[{{Sezestre} {et~al.}(2017){Sezestre}, {Augereau}, {Boccaletti}, \&
  {Th{\'e}bault}}]{sez17}
{Sezestre}, {\'E}., {Augereau}, J.-C., {Boccaletti}, A., \& {Th{\'e}bault}, P.
  2017,
  \href{http://dx.doi.org/10.1051/0004-6361/201731061}{\JournalTitle{\aap},
  607, A65}

\bibitem[{{Stobie} \& {Ferro}(2006)}]{idp3}
{Stobie}, E., \& {Ferro}, A. 2006, in Astronomical Society of the Pacific
  Conference Series, Vol. 351, Astronomical Data Analysis Software and Systems
  XV, ed. C.~{Gabriel}, C.~{Arviset}, D.~{Ponz}, \& S.~{Enrique}, 540

\bibitem[{{Strubbe} \& {Chiang}(2006)}]{strubbe06}
{Strubbe}, L.~E., \& {Chiang}, E.~I. 2006,
  \href{http://dx.doi.org/10.1086/505736}{\JournalTitle{\apj}, 648, 652}

\bibitem[{{Townsend} \& {Owocki}(2005)}]{rrm2}
{Townsend}, R.~H.~D., \& {Owocki}, S.~P. 2005,
  \href{http://dx.doi.org/10.1111/j.1365-2966.2005.08642.x}{\JournalTitle{\mnras},
  357, 251}

\bibitem[{{Townsend} {et~al.}(2005){Townsend}, {Owocki}, \& {Groote}}]{rrm1}
{Townsend}, R.~H.~D., {Owocki}, S.~P., \& {Groote}, D. 2005,
  \href{http://dx.doi.org/10.1086/462413}{\JournalTitle{\apjl}, 630, L81}

\bibitem[{{Vidotto} {et~al.}(2014){Vidotto}, {Jardine}, {Morin}, {Donati},
  {Opher}, \& {Gombosi}}]{vidotto2014}
{Vidotto}, A.~A., {Jardine}, M., {Morin}, J., {et~al.} 2014,
  \href{http://dx.doi.org/10.1093/mnras/stt2265}{\JournalTitle{\mnras}, 438,
  1162}

\bibitem[{{Walkowicz} {et~al.}(2013){Walkowicz}, {Basri}, \& {Valenti}}]{w2013}
{Walkowicz}, L.~M., {Basri}, G., \& {Valenti}, J.~A. 2013,
  \href{http://dx.doi.org/10.1088/0067-0049/205/2/17}{\JournalTitle{\apjs},
  205, 17}

\bibitem[{{Wang} {et~al.}(2015){Wang}, {Graham}, {Pueyo}, {Nielsen},
  {Millar-Blanchaer}, {De Rosa}, {Kalas}, {Ammons}, {Bulger}, {Cardwell},
  {Chen}, {Chiang}, {Chilcote}, {Doyon}, {Draper}, {Duch{\^e}ne}, {Esposito},
  {Fitzgerald}, {Goodsell}, {Greenbaum}, {Hartung}, {Hibon}, {Hinkley}, {Hung},
  {Ingraham}, {Larkin}, {Macintosh}, {Maire}, {Marchis}, {Marois}, {Matthews},
  {Morzinski}, {Oppenheimer}, {Patience}, {Perrin}, {Rajan}, {Rantakyr{\"o}},
  {Sadakuni}, {Serio}, {Sivaramakrishnan}, {Soummer}, {Thomas}, {Ward-Duong},
  {Wiktorowicz}, \& {Wolff}}]{wang2015}
{Wang}, J.~J., {Graham}, J.~R., {Pueyo}, L., {et~al.} 2015,
  \href{http://dx.doi.org/10.1088/2041-8205/811/2/L19}{\JournalTitle{\apjl},
  811, L19}

\bibitem[{{Watson} {et~al.}(2011){Watson}, {Littlefair}, {Diamond}, {Collier
  Cameron}, {Fitzsimmons}, {Simpson}, {Moulds}, \& {Pollacco}}]{watson2011}
{Watson}, C.~A., {Littlefair}, S.~P., {Diamond}, C., {et~al.} 2011,
  \href{http://dx.doi.org/10.1111/j.1745-3933.2011.01036.x}{\JournalTitle{\mnras},
  413, L71}

\bibitem[{{Wood} {et~al.}(2005){Wood}, {Redfield}, {Linsky}, {M{\"u}ller}, \&
  {Zank}}]{wood2005}
{Wood}, B.~E., {Redfield}, S., {Linsky}, J.~L., {M{\"u}ller}, H.-R., \& {Zank},
  G.~P. 2005, \href{http://dx.doi.org/10.1086/430523}{\JournalTitle{\apjs},
  159, 118}

\end{thebibliography}

\end{document}